\documentstyle[aps,12pt,preprint,tighten]{revtex}
\begin{document}

\draft

\title{\rightline{{\tt (February 2000)}}
\rightline{{\tt UM-P-006/2000}}
\rightline{{\tt RCHEP-001/2000}}
\ \\
Mirror matter and heavy singlet neutrino oscillations in the early universe}
\author{Nicole F. Bell}
\address{School of Physics, Research Centre for High Energy Physics\\
The University of Melbourne, Victoria 3010 Australia\\
(n.bell@physics.unimelb.edu.au)}
\maketitle

\begin{abstract}
We investigate the mixing of heavy gauge singlet neutrinos in a 
mirror matter model employing the seesaw mechanism.  The parameter 
constraints that must be satisfied to prevent the overproduction of 
mirror matter in the early universe are deduced.  We find that no 
fine tuning in the heavy neutral fermion sector is required for this mirror 
matter model to satisfy cosmological constraints.  
Baryogenesis scenarios are briefly discussed in the context of the mirror model.
\end{abstract}

\section{Introduction}

The idea that every ordinary particle is related to a mirror partner has been 
discussed in Refs.\cite{leeyang,okun,flv,nu,kst,mohapatra,blinnikov,silagadze}.
In the Exact Parity Model (EPM) of Ref.\cite{flv}, mirror matter is postulated 
as means by which to retain invariance under all Improper Lorentz transformations, 
whereby each ordinary particle 
and its mirror partner are related via a non-standard parity symmetry.  In 
this symmetric mirror-matter model, the parity symmetry dictates that the 
particle interactions in the mirror world are of precisely the same strength 
as those in the ordinary world.  
On the other hand, the authors of Ref.\cite{mohapatra} consider non-symmetric 
mirror matter models in which the parity symmetry is spontaneously broken.

In the context of the EPM the ordinary and mirror sectors interact only via 
mixing between the ordinary and mirror neutrinos, Higgs bosons, and neutral 
gauge bosons, and gravitationally.
The mixing between ordinary and mirror neutrinos is of great interest for 
explaining the observed neutrino anomalies \cite{flv,nu}, both because parity 
symmetry is a theoretically natural way to obtain the maximal mixing indicated by 
the Superkamiokande (SK) experiment 
\footnote{Note that results from SK \cite{kajita} now appear to  
disfavour the $\nu_{\mu} \rightarrow \nu_s$ solution to the atmospheric 
anomaly.  This is as yet preliminary and does not alter our discussion of heavy 
singlet neutrino dynamics.  We eagerly await further results and analysis from 
SK on this matter.} 
\cite{superk}, and because the mirror 
neutrinos can fulfill the need for light ``sterile'' neutrinos. 
The implications of a mirror sector have also been investigated in a wide range 
of astrophysical and cosmological contexts, see 
Refs.\cite{blinnikov,machos,planets,bh,grb,mt,asymm}.

We are concerned here with the possible oscillations amongst heavy gauge singlet 
neutrinos in the early universe.  These are heavy right handed neutrinos (and left 
handed mirror neutrinos) which 
are employed in seesaw models of neutrino mass to suppress the masses of the light 
neutrinos.  The essential point is that all neutrinos in the EPM (both the ordinary 
light neutrinos and the heavy gauge singlet neutrinos) are maximally mixed with a 
mirror partner.  This is potentially dangerous since large amplitude oscillations 
between the active and mirror neutrinos in the early universe may serve to 
equilibrate the mirror neutrinos, even if there was initially no mirror matter 
present.  This of course would then be in violation of constraints on the number of 
light particle species in thermal equilibrium at the time of big bang 
nucleosynthesis (BBN).

In the case of oscillations between the three light neutrinos species and their 
mirror partners, it turns out that the naive expectation that the mirror neutrinos 
are all equilibrated for $\Delta m^2 \agt 10^{-6} \text{eV}^2$ can be avoided in 
a natural way \cite{asymm}.  Given an initial 
baryon asymmetry, there exists a large region in of parameter space for which small 
angle (inter-generational) ordinary-mirror mixing can amplify the asymmetry 
between the number of neutrinos and anti-neutrinos to be many orders of 
magnitude larger than the baryon asymmetry.  This creates an effective potential 
which suppresses the maximal mixing between the ordinary and mirror neutrinos, and 
hence prevents the mirror species from being brought into equilibrium. 

However, we have still the heavy singlet neutrinos to worry about. 
We will be concerned with temperatures greatly exceeding that of the electroweak 
phase transition, where an initial lepton or baryon asymmetry may or may not have 
been present depending on the specific dynamics in the early universe (for example 
an asymmetry may have been created by an Affleck Dine mechanism \cite{ad}).
If a large asymmetry in the light lepton sector existed, it would contribute to 
the effective potential \cite{mu} which suppresses the oscillations of the 
heavy neutrinos.
However, unless we were to 
introduce gauge bosons that couple to the singlet neutrinos,
we would expect no direct feedback effect and hence no 
exponential growth of lepton asymmetry via oscillations (as can occur in the 
light neutrino sector at low temperatures), since the potential would not depend 
upon the asymmetry in the number of heavy neutrinos themselves. 
Hence, we will examine the case where no large asymmetries are responsible 
for suppressing oscillations.

Note that singlet neutrinos are of great interest with regard to 
producing the observed baryon asymmetry - either via CP violating 
out-of-equilibrium decays \cite{fy} or via oscillations \cite{ars}. 
We shall examine the interactions of the gauge singlet neutrinos and hence 
determine the parameter constraints that must be satisfied to render the EPM 
consistent with cosmology.

\section{Mirror neutrinos}

We begin by outlining the neutrino sector of the EPM.
The most natural choice to investigate is a seesaw model as was considered 
in \cite{nu}.  We reproduce here the relevant features. 

The neutrino fields consist of $\nu_L$ and $N_R$, and their mirror partners 
$\nu'_R$ and $N'_L$.  The full gauge symmetry of the theory is
\begin{equation}
\text{SU(3)}_1 \otimes \text{SU(2)}_1 \otimes \text{U(1)}_1 \otimes \text{SU(3)}_2 
\otimes \text{SU(2)}_2 \otimes \text{U(1)}_2,
\end{equation}
that is, the Standard Model gauge group squared.
The  $\nu_L$ belongs to an $\text{SU(2)}_1$ doublet and its mirror partner $\nu_R'$ 
belongs to an $\text{SU(2)}_2$ doublet,
\begin{equation}
f_L = ( \nu_L \;\; e_L)^{\text{T}}, \;\;\;\;\; 
f_R' = ( \nu'_R \;\; e_R')^{\text{T}},
\end{equation}
while both the $N_R$ and $N'_L$ are gauge singlets. 
The Yukawa Lagrangian which includes the most general renormalisable terms 
allowed by the gauge and parity symmetries is   
\begin{eqnarray}
\label{yuk}
{\cal L}_{\text{Yuk}} &=& \lambda_1 [\overline{f}_L \tilde{\phi_1} N_R 
+ \overline{f}'_R \tilde{\phi_2} N_L']
+ \lambda_2 [\overline{f}_L \tilde{\phi_1} (N_L')^C 
+ \overline{f}'_R \tilde{\phi_2} (N_R)^C ] \\ \nonumber
&+& M_1 [ \overline{N}_R (N_R)^C +  \overline{N}_L' (N_L')^C  ]
+ M_2 \overline{N}_R N_L' + \text{H.c} .
\end{eqnarray}
The ordinary $\text{SU(2)}_1$ doublet Higgs field is  
$\phi_1=(\phi^+ \; \phi^0)^{\text{T}}$, and we employ the notation 
$\tilde{\phi}=i\sigma_2\phi^*$, where $\sigma_2$ denotes a Pauli matrix.
The $\phi_2$ Higgs field, the parity partner of $\phi_1$, is correspondingly a
doublet under $\text{SU(2)}_2$.

Below the temperature at which the electroweak phase transition occurs, the  
$\lambda_1$ terms produce Dirac 
masses for the $\nu$'s and $\nu'$'s, while the $\lambda_2$ mass terms lead to mixing 
between the ordinary and mirror neutrinos.  We shall denote these masses $m_1$ and 
$m_2$ respectively, where $m_{1,2}=\lambda_{1,2} v$, with $v \simeq246 \text{GeV}$ being 
the Higgs vacuum expectation value (VEV).  However, we shall be mainly interested in the 
symmetric phase, where the VEV vanishes and light neutrinos 
are massless.  Note that in this case there is no mass mixing between the heavy 
and light sector.  The $M_1$ and $M_2$ terms are bare masses, which we shall assume 
are much greater than the electroweak scale masses.  
The $M_2$ terms mix ordinary and mirror matter.

We shall assume that intergenerational mixing is small and consider a single 
generation model \cite{flv}.
In the parity diagonal basis, defined by the states 
$( \nu_L^+ \; \nu_L^-  \; (N_R^-)^C \; (N_R^+)^C )^{\text{T}}$ where,
\begin{equation}
\nu_L^{\pm} = \frac{1}{\sqrt{2}}(\nu_L \pm (\nu'_R)^C), \;\;\;\;\;
N_R^{\pm} = \frac{1}{\sqrt{2}}(N_R \pm (N'_L)^C), 
\end{equation}
the mass matrix takes the form
\begin{equation}
\cal{M} =\left( \begin{array}{cccc}
0 & 0   & 0 & m_+ \\
0 & 0  & m_- & 0 \\
0 & m_-  & M_-   & 0 \\
m_+ & 0 & 0 & M_+
\end{array} \right),
\end{equation}
where $M_{\pm}=M_1 \pm M_2$ and $m_{\pm}=m_1 \pm m_2$. 
The light neutrino mass  (and also parity) eigenstates are approximately
$\nu_L^{\pm}$ with seesaw suppressed Majorana masses given by 
\begin{equation}
m_a =m_+^2/M_+, \hspace{1cm} m_b=m_-^2/M_-,  
\end{equation}
while the heavy  eigenstates $N_R^{\pm}$ have masses given by
\begin{equation}
M_+, \hspace{1cm} M_-.  
\end{equation}
It is the existence of the $\lambda_2$ terms, no matter how small, that forces the 
mixing between an ordinary neutrino and its mirror partner to be non-zero, and hence 
maximal because the mass and parity eigenstates must coincide.  
Likewise, any nonzero value of $M_2$ requires that the heavy singlet neutrinos are 
maximally mixed.

So we see that the ordinary light neutrinos are coupled to ordinary singlets $N_R$ via 
$\lambda_1$ and to mirror singlets $N_L'$ via $\lambda_2$.  It would seem natural and, 
for the heavier generations, it shall turn out to be necessary, that 
$\lambda_2$ is smaller than $\lambda_1$.
Note that for $\lambda_1 \gg \lambda_2$, $m_{\pm} \simeq m_1$.
We shall also assume that $M_2 < M_1$ so that $M^+$ and $M^-$ have masses 
of approximately the same order of magnitude, which we denote as $M$. 
Hence the idea is that for each generation, we have a pair of almost degenerate 
mass eigenstates.  
Considering only one generation, the light and heavy sector neutrino masses are 
related according to \footnote{ For a three generation model, the parameter 
$m_{\nu}$ coincides with the light neutrino mass for a given generation only if 
we assume small intergenerational mixing.}
\begin{equation}
\label{ml}
m_{\nu} \simeq \frac{m_1^2}{M} = \frac{(\lambda_1 v)^2}{M}.
\end{equation}

\section{The early universe}
\label{early}

Firstly, we shall be interested in the interactions which equilibrate the ordinary 
right-handed (and possibly also the mirror left-handed) singlet neutrinos.
Let us assume that at high temperatures we have no singlet neutrinos, $N_R$ and 
$N_L'$, in the cosmological plasma.  They 
will then be generated by scattering processes, according to the Yukawa couplings to 
the light neutrinos and Higgs bosons.  If the 
rates for these processes exceed the expansion rate $H \simeq [T(GeV)]^2/10^{18}$, 
the $N$'s will achieve thermal equilibrium.  
We shall be working in the regime where the temperature is much larger than 
the mass of the Higgs bosons, so that the rates are approximately independent 
of these parameters.

The $N$'s will decay producing both ordinary and mirror light neutrinos and 
Higgs bosons via processes such as 
\begin{eqnarray}
\label{gamma_d}
N \rightarrow \nu + \phi_1, \\ \nonumber 
N \rightarrow \nu' + \phi_2,
\end{eqnarray}
with a total decay rate given by 
\begin{eqnarray}
\label{decay}
\Gamma_D = \Gamma_{D_1} + \Gamma_{D_2}
& \simeq & \frac{(\lambda_1^2 + \lambda_2^2)}{16 \pi} M, 
\;\;\;\;\;\;\;   T \alt M, \nonumber \\
& \simeq & \frac{(\lambda_1^2 + \lambda_2^2)}{16 \pi} M \frac{M}{T}, 
\;\;\;\;\;  T \gg M.
\end{eqnarray}

The scattering rate for processes such as $\nu q \rightarrow N q$ (via Higgs 
exchange), for $T \gg M$ is given by \cite{luty}
\begin{equation}
\Gamma_i = \frac{9}{64 \pi^3} \lambda_t^2 \lambda_i^2 T,
\end{equation}
where $i=1,2$ for ordinary and mirror singlets respectively, and $\lambda_t$ is the 
Yukawa coupling constant for the top quark.  (Processes involving the top quark 
dominate the scattering rate since the Yukawa coupling to the Higgs boson is much 
larger than for the other quarks and leptons.) 
In addition to scattering, singlet neutrinos may be produced by inverse decay 
processes, with rates given by eq.(\ref{gamma_d}), which however are small compared 
to scattering.
For $T \ll M$, production of heavy neutrinos is kinematically suppressed.

The mirror singlets must not be equilibrated via scattering, as this would lead to 
the whole mirror sector being equilibrated, so we require $\Gamma_2 < H$ 
\footnote{Note that when the N's are produced, an effective potential suppresses 
mixing so that the flavour eigenstates and the matter-affected mass 
eigenstates almost coincide.}.  
This condition is most restrictive for $T=M$, leading to 
\begin{equation}
\frac{\lambda_2^2}{M (\text{GeV})} \alt 2 \times 10^{-16},
\end{equation}
or, using eq.(\ref{ml})
\begin{equation}
\label{lambda_2}
\frac{\lambda_2}{\lambda_1} < \frac{0.1}{\sqrt{m_{\nu}(eV)}}. 
\end{equation}

We may determine if the ordinary singlets are thermalised, in terms of the size 
of the light masses.  If the $N_R$'s are thermalised, we have that
\begin{equation}
\label{lambda_1}
\frac{\lambda_1^2}{M (\text{GeV})} \agt 2 \times 10^{-16},
\end{equation}
and hence 
\begin{equation}
\label{m}
m_{\nu} \agt 10^{-2} \text{eV}.
\end{equation}
If a light neutrino mass satisfies this bound then the corresponding heavy singlet 
$N_R$ will have been thermalised.
(Of course, once we go to a three generation model we need to worry about how 
the bases in which  $m$ and $M$ are diagonalised are related).
The atmospheric neutrino anomaly with the Superkamiokande determined mass 
squared difference of  $\delta m^2  \sim 10^{-2}-10^{-3}$ suggests the $\nu_{\mu}$ 
mass would satisfy eq.(\ref{m}).  We could reasonably expect the $\nu_{\tau}$ to 
be yet heavier, and in fact the EPM requires the $\nu_{\tau}$ mass to be 
$m_{\nu_{\tau}} \sim \text{eV}$ \cite{asymm} if the atmospheric problem is solved 
by $\nu_{\mu} \rightarrow \nu_{\mu}'$ oscillations. 
So we know that at least one of the ordinary singlet neutrino flavours was brought into 
thermal equilibrium at some stage in the early universe. We must then make sure 
that the amount of mirror matter produced, either by decay or oscillation of 
the ordinary singlets is sufficiently small. 

As the dynamics of the singlet neutrinos with Yukawa coupling constants large 
enough that they were thermalised is qualitatively different to those with 
smaller Yukawa coupling constants, we shall look at these two cases separately.

\subsection{Thermalised singlet neutrinos}
\label{heavygen}

For temperatures $T \agt M$, the scattering rate is greater than the decay rate, so 
that enforcing, as one must do, eq.(\ref{lambda_2}) automatically ensures that 
singlet neutrino decays do not over produce mirror particles.
For light neutrino masses which satisfy eq.(\ref{m}), the corresponding heavy singlet 
decay rate will become equal to the expansion rate while the $N$'s are still 
relativistic.  Hence the singlets will decay away before becoming highly 
non-relativistic, and the decay will not lead to appreciable reheating.

While the singlets will have all decayed before they become very non-relativistic, 
we shall nonetheless be concerned with oscillations that occur when the 
ultra-relativistic limit may not hold.  For this reason, we shall give expressions 
for both the extreme relativistic and non-relativistic limits, and ensure both sets 
of bounds may safely be satisfied.

We wish to constrain the parameters of the model such that the maximal 
ordinary-mirror singlet oscillations are suppressed.  Essentially, we need for 
the effective potential to be sufficiently larger than the energy difference 
between the vacuum mass eigenstates.
The effective potential for ordinary singlets in a background medium 
which contains only ordinary matter, for temperatures much larger than the 
mass of the Higgs boson, is given by \cite{weldon} 
\footnote{Note that the calculation of the finite temperature 
effective masses in \cite{weldon} is performed in the high 
temperature regime $T \gg E,p$.  For the regime we are interested 
in, namely $E \sim T$ there may be an additional coefficient of 
order unity in the potential (\ref{v}), which however, should 
not greatly alter our result.  See Ref.\cite{ft} for a discussion 
of the fermion dispersion relations without this approximation.},
\begin{eqnarray}
\label{v}
V(\lambda_1) & \simeq & \frac{1}{8} \lambda_1^2T, \;\;\;\;\; T \gg M  \nonumber \\
 & \simeq & \frac{1}{8} \lambda_1^2T \frac{T}{M}, \;\;\;\;\; T \ll M
\end{eqnarray}
while the potential for mirror singlets  (in an ordinary matter background) is 
obtained by replacing $\lambda_1$ with $\lambda_2$.  It is the difference in 
effective potentials that is related to the oscillation parameters, 
\begin{equation}
V = V(\lambda_1) -V(\lambda_2) \simeq V(\lambda_1).
\end{equation}
To effectively suppress oscillation, this must be bigger than the energy difference 
between the two mass eigenstates, $\omega_1-\omega_2 \simeq \Delta M^2/(2p)$ 
(relativistic) or $\omega_1-\omega_2 \simeq \Delta M$ (nonrelativistic). 
To achieve this we will clearly require a sufficiently large $\lambda_1$ and/or 
a sufficiently small mass difference.

We shall determine if the necessary constraints are consistent with the 
light neutrino masses and mass-squared differences applicable to the 
resolution of the neutrino anomalies. Note that we make the approximation of 
a mono-energetic rather than a thermally distributed spectrum of neutrino energies, 
which means for  $T \gg M$ we take the neutrino momentum to be given by its thermal 
average $p \sim T$.

The rate at which mirror singlets are produced is given approximately by 
\cite{early,asymm}
\begin{equation}
\label{osc}
\Gamma (N \rightarrow N') \simeq \frac{1}{2} D \sin^2 2\theta_m,
\end{equation}
where
\begin{equation}
D = \frac{1}{2}[ \Gamma_{\text{scatt}}(N_R)+ \Gamma_{\text{scatt}}(N_L')] 
\simeq  \frac{1}{2}\Gamma_{\text{scatt}}(N_R),
\end{equation}
and the matter mixing angle $\theta_m$ is determined by
\begin{equation}
\label{thetam}
\sin^2 2\theta_m = \frac{\sin^2 2\theta_0}{\sin^2 2\theta_0+ [(2p/\Delta M^2) V]^2 },
\end{equation}  
with $\theta_0$ being the vacuum mixing angle, such that 
$\sin^2 2\theta_0 =1$ for maximal mixing.  
In eq.(\ref{thetam}) we have assumed that the $N$'s are relativistic.  In the 
nonrelativistic limit the factor $2p/\Delta M^2$ should be replaced by $1/\Delta M$.

In eq.(\ref{osc}), we have assumed the oscillations are adiabatic, which  
holds provided  
\begin{equation}
\gamma \equiv \frac{d\theta_m}{dt}/\frac{\Delta M^2}{2p} \ll 1.
\end{equation}
We find that away from resonance (for the period of time in question the effective 
potential must be large enough for the neutrinos to be away from resonance),
\begin{equation}
\gamma \simeq \frac{8}{\lambda_1^2} \left( \frac{T}{10^{18}\text{GeV}} \right).
\end{equation}
At the temperature where the $N$'s are first thermalised
\begin{equation}
\gamma (T^{\text{thermalise}}) \simeq 0.01,
\end{equation}
and since $\gamma$ decreases with time, the oscillations are always adiabatic.

To estimate the constraints that must be satisfied
to prevent the mirror N's from being overproduced, we shall use the condition
\begin{equation}
\label{rate}
\Gamma (N \rightarrow N') < H.
\end{equation}
For relativistic $N$'s this implies 
\begin{equation}
\frac{\Delta M^2}{M^2}  < \frac{0.1 \lambda_1^2}{\sqrt{m_{\nu}(\text{eV})}}
\left( \frac{T}{M} \right)^{5/2},
\end{equation}
where we have used eq.(\ref{m}). Given that this will be most stringent for 
low temperatures we shall set $T \sim M$, giving the bound 
\begin{equation}
\label{rel}
\frac{\Delta M^2}{M^2}  < \frac{0.1 \lambda_1^2}{\sqrt{m_{\nu}(\text{eV})}}.
\end{equation}

In the non-relativistic limit we would have a similar condition
\begin{equation}
\label{nonrel}
\frac{\Delta M}{M}  < \frac{0.05 \lambda_1^2}{\sqrt{m_{\nu}(\text{eV})}}
\left( \frac{T}{M} \right)^{5/2} 
\left( \frac{\Gamma_{\text{scatt}}^{rel}}{\Gamma_{\text{scatt}}^{non-rel}} \right)^{1/2}.
\end{equation} 
However, we need only for this to hold until the $N_R$'s decouple, that is, when 
they stop being replenished by scattering.  This happens at about the time they 
become non-relativistic anyway, so that $(T/M)$ in eq.(\ref{nonrel}) would be at 
least of order $1/10$, meaning that the constraint given eq.(\ref{nonrel}) is a 
similar condition to eq.(\ref{rel}). 
Note that for $M_2 < M_1$, $\Delta M^2/M^2 \simeq 4M_2/M_1 \simeq  2\Delta M/M$.

The criterion given in eq.(\ref{rate}) is useful to provide a rough estimate as 
to when a species is thermalised by a certain process.  More accurate 
results could be obtained with detailed numerical work, though for our 
purposes approximate expressions will suffice since we are more 
interested in the the rough size of the bounds and whether any fine 
tuning of parameters would be require to satisfy them.
The conditions (\ref{rel},\ref{nonrel}) are not particularly restrictive, and leave 
plenty of scope to obtain the light neutrino mass squared differences
($\delta m^2_{\text{light}}=[m_+^2/M_+]^2 - [m_-^2/M_-]^2$) suggested by the 
neutrino oscillation experiments.

So we conclude that the bounds on the couplings that mix matter and 
mirror matter are not very severe, and no unnatural fine tuning of 
parameters is required to achieve consistency with cosmological constraints.  
This may be compared with the light neutrino sector where, similarly, 
no fine tuning of parameters needs to be done, due to the mechanism of 
asymmetry generation \cite{asymm}.

Finally, we wish to comment on possible scenarios for the production of a baryon 
and a corresponding mirror baryon asymmetry within the context of the mirror model.

We envisage a variation on the scenario proposed in ref.\cite{ars}, in which 
neutrino oscillations create a lepton asymmetry which is then reprocessed into 
a baryon asymmetry by electroweak sphaleron transitions.  In ref.\cite{ars}, 
CP violating oscillations between three heavy singlet neutrino species distribute 
the total lepton number (which satisfies $L^{\text{total}}=0$) unevenly between 
the three species.  Due to a hierarchy in the Yukawa coupling constants, one of 
the three singlets does not communicate this asymmetry to the usual leptons before 
the time when sphaleron transitions freeze out, resulting in a non-zero baryon 
asymmetry.

In the mirror model, we similarly wish to consider CP violating oscillations, but 
in this case we may obtain $L \ne 0$ and $L' \ne 0$ satisfying $L+(-L')=0$
\footnote{In the EPM, oscillations amongst ordinary and mirror neutrinos conserve $L-L'$, 
rather than $L+L'$, because oscillations actually interchange neutrinos with mirror 
antineutrinos, for example $\nu_L \leftrightarrow (\nu'^C)_L$}. 
The heavy neutrino asymmetries are communicated to the light leptons, and the
lepton and mirror lepton asymmetries are simply reprocessed into baryon and 
mirror baryon asymmetries respectively, with $B'=B$.
Note that although the temperature of any thermalised mirror matter must be 
smaller than the corresponding temperature of matter, the bound is quite weak, with 
$T'<0.5 T$ being sufficient to satisfy BBN constraints.
Hence it is possible that the time at which mirror sphalerons freeze out is not much 
earlier than for the ordinary sphalerons.
Having ordinary and mirror baryon numbers of the same size is a nice feature, since 
it implies that even though at early times ordinary matter dominated, 
the amount of mirror matter in the universe today is the same as the 
amount of ordinary matter, thereby having interesting consequences for dark matter 
\cite{machos,planets}.

A detailed discussion of the parameter space where this scenario is viable may be 
found in ref.\cite{ars}.
Basically, somewhat small singlet neutrino masses are required, $M<100 \text{GeV}$, 
to avoid washout of the asymmetry through lepton number violation arising from the 
Majorana mass.  Additionally, the Yukawa coupling constants must be large enough for 
the asymmetry to be communicated to the usual leptons before the spheralon 
transitions freeze out, $\lambda_1^2 > 10^{-14}$.  These bounds are consistent with 
phenomenologically relevant neutrino masses.

\subsection{The lightest singlet neutrino}
\label{lightgen}

We shall now discuss the case where the light neutrinos and mirror neutrinos 
have masses 
\begin{equation}
m_a, m_b \alt 10^{-2} \text{eV},
\end{equation}
which is quite likely to apply for electron-flavour neutrinos.  In this case the 
singlet neutrino parameters are not subject to the bounds in subsection \ref{heavygen}, 
as both 
$\lambda_1$ and $\lambda_2$ are small enough that the scattering production rates 
for singlets are always smaller than the expansion rate.
There are two possibilities, either $N_e$ and $N_e'$ are never populated, or 
the $N_e$ states (but not $N_e'$) are populated by processes operating at high 
temperatures for which the physics involved is as yet uncertain.

If the latter occurred, we may appeal to the leptogenesis scenario of out-of-equilibrium 
CP violating decays \cite{fy,luty} to generate the baryon asymmetry of the universe.  
Note that baryogenesis via out-of-equilibrium decay requires singlet neutrino masses 
much larger than the electroweak scale, which is a completely different region of 
parameter space to that required for baryogenesis via neutrino oscillation as 
discussed above. 

For leptogenesis to be successful, the $N$'s must be out-of-equilibrium when they decay. 
However, if at the time of decay they were extremely non-relativistic, 
significant reheating 
of both ordinary and mirror particles species could result, which 
in addition to diluting the final value of any asymmetry generated, could reduce the 
temperature difference between the ordinary and the mirror sector particle species.

The $N_e$ will decay-out-of equilibrium, but without causing appreciable
reheating, if the masses of the light neutrinos are in the range
\begin{equation}
10^{-6} \text{eV} \alt m_a, m_b \alt 3 \times 10^{-3} \text{eV}.
\end{equation}
Since $N_e$ may decay into both ordinary and mirror matter, we can obtain both $L$ and $L'$
asymmetries:
\begin{eqnarray}
L \propto \epsilon 
&=& \frac{\Gamma(N \rightarrow \nu \phi) - \Gamma(N \rightarrow \bar{\nu} \phi^{\dagger})}
{\Gamma_D^{\text{total}}} \nonumber \\
L' \propto \epsilon' 
&=& \frac{\Gamma(N \rightarrow \nu' \phi') - \Gamma(N \rightarrow \bar{\nu}' (\phi')^{\dagger})}
{\Gamma_D^{\text{total}}}
\end{eqnarray}

Estimating the size of $\epsilon$ and $\epsilon'$ would require assumptions about the size 
of CP violating phases in the Yukawa coupling constants and the singlet neutrino mass matrix.
However, since at tree level, the decay rates $\Gamma(N \rightarrow \nu \phi)$ and 
$\Gamma(N \rightarrow \nu' \phi')$ can be of the same order of magnitude 
(because $\lambda_2 \sim \lambda_1$ is permitted), it would seem plausible that $L'$ could 
be as large as $L$.

\section{Conclusion}

We have examined bounds on the parameters in the gauge singlet neutrino sector 
of the EPM to ensure mirror matter is not overproduced in the early universe, 
and conclude that a seesaw mechanism can be implemented in the EPM with no 
unnatural fine-tuning of the various masses and coupling constants required. 

Heavy singlet neutrinos may allow us to account for the baryon asymmetry of the 
universe using a leptogenesis scenario, either via CP violating oscillations or 
out of equilibrium decays.  In both cases, even though the total energy density 
of mirror matter is constrained to be smaller than the energy density of ordinary 
matter during the radiation dominated epoch of the universe, it is conceivable that 
baryon and mirror baryon asymmetries of comparable size could result.

\acknowledgments{ NFB acknowledges discussions with R. R. Volkas and thanks him 
for reading drafts of this paper.  Thanks are also due to N. Witte for a discussion 
and to E. Kh. Akhmedov for email correspondence.
This work was initiated at the Low Energy Neutrino Workshop held at the Institute 
for Nuclear Theory at the University of Washington in Seattle.  The author thanks 
A. Balantekin and W. Haxton for their hospitality.
NFB is supported by the Commonwealth of Australia and The University
of Melbourne.}

\end{document}